\documentclass[aps,prd,preprint,eqsecnum,tightenlines,floats,
groupedaddress,showpacs]{revtex4-1}
\usepackage{amssymb}
\usepackage{hyperref}
\usepackage{graphicx}
\usepackage{subfigure}

\begin{document}

\date{\today}

\title{Effects of a caustic ring of dark matter on\\
the distribution of stars and interstellar gas}

\author{Sankha S. Chakrabarty and Pierre Sikivie}

\affiliation{Department of Physics, University of Florida, Gainesville, Florida 32611, USA}

\begin{abstract}

Caustic rings of dark matter with \textit{tricusp} cross section were predicted to lie in the galactic disk. Their radii increase on cosmological time scales at a rate of order $1$ kpc/Gyr. When a caustic ring passes through the orbit of a star, the orbit is strongly perturbed. We find that a star moving in a nearly circular orbit is first attracted towards the caustic ring, then moves with and oscillates about the caustic for approximately $1$ Gyr before returning to its original orbit. As a result, a stellar overdensity forms around the caustic ring. We predict such overdensities to be of order $120$\% near the second caustic ring where the Monoceros Ring is observed and of order $45$, $30$ and $15$\% near the third, fourth and fifth caustic rings, respectively. We show that the associated bulk velocities of the stars near the caustic rings are less than a few km/s. We also determine the density profile of interstellar gas near the fifth caustic ring assuming it is in thermal equilibrium in the gravitational potential of the caustic and of the gas itself.  

\end{abstract}

\maketitle

\section{\label{I}Introduction}

From various observations, it is known that $23\%$ of the energy density of the Universe is made of cold dark matter (CDM) \cite{Bertone, *Weinberg, *Kolb}. Axions are one of the widely accepted  cold dark matter candidates \cite{Sikivie08, *Marsh16}. The QCD axion was originally proposed as a solution to the strong \textit{CP} problem in the standard model \cite{Peccei77, *Peccei2, *Weinberg78, *Wilczek78, Kim79, *Shifman80, *Zhitnitskii80, *Dine81}. It became a cold dark matter candidate when it was shown that QCD axions are produced abundantly in the early universe with very small velocity dispersion \cite{Preskill83, *Abbott83, *Dine83, *Ipser83}.

\begin{figure*}
\includegraphics[scale=0.5]{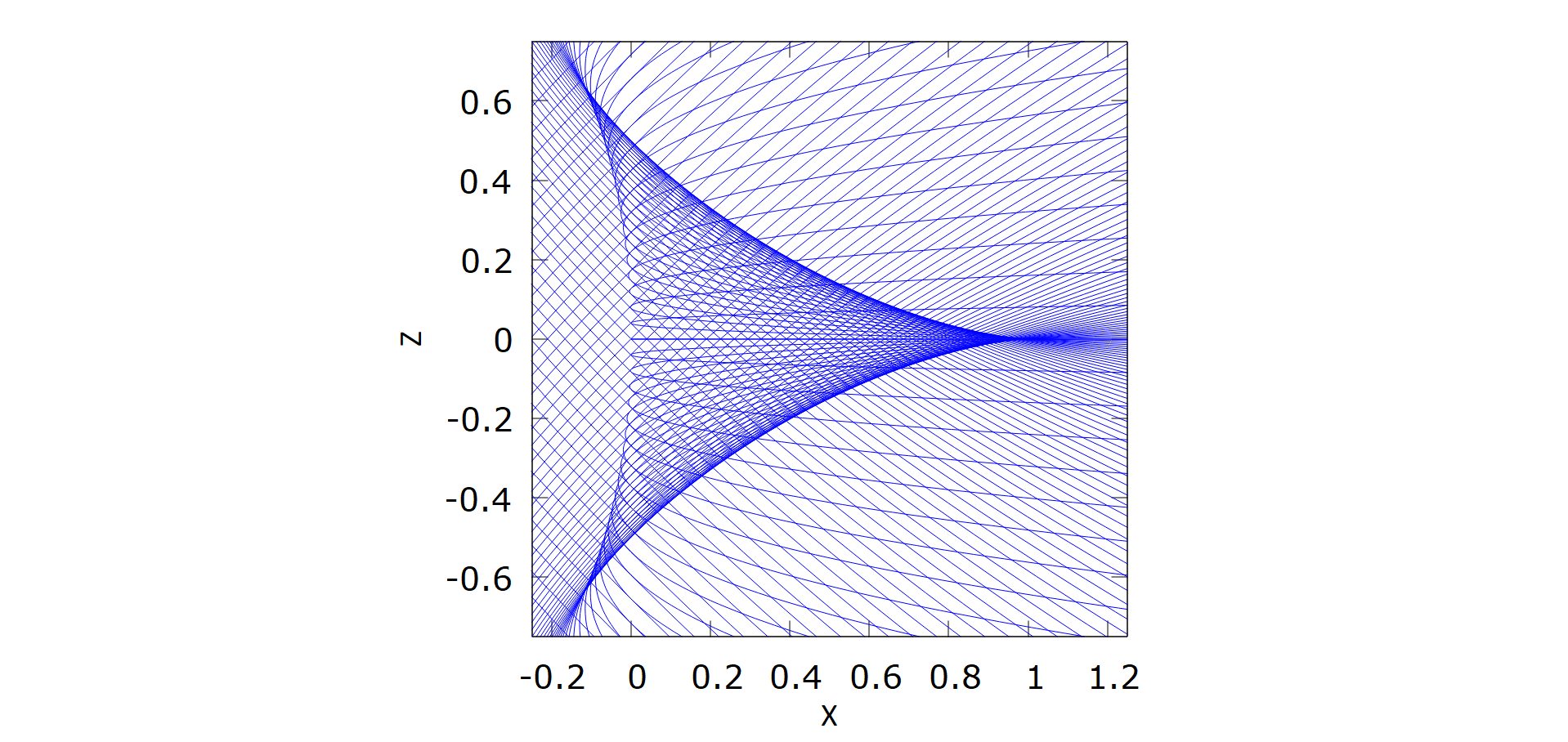}
\caption{\label{tricusp_flows}Flows of cold dark matter particles in a $\rho$-$z$ cross section of a caustic ring.  $X$ and $Z$ are rescaled coordinates $X = \frac{\rho - a}{p}$ and $Z = \frac{z}{p}$.}
\includegraphics[scale=0.4]{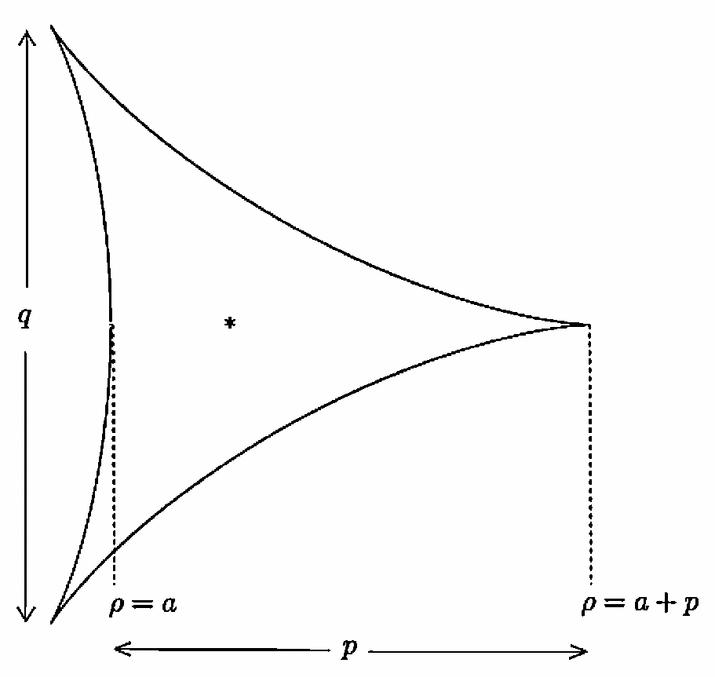}
\caption{\label{tricusp}The envelope of the trajectories shown in Fig.~\ref{tricusp_flows}. The shape in this figure is usually called a tricusp. The radius $a$ and the transverse sizes $(p, q)$ are indicated in the figure. The figure is taken from Ref.~\cite{Duffy08}. The radii $a_n$ and sizes $p_n$ of different caustic rings are given in Ref.~\cite{Duffy08}. For the caustic ring nearest to us ($n=5$), $a_5 \approx 8.28$ kpc, $p_5 \approx 150$ pc and $q_5 \approx 200$ pc.}
\end{figure*}

Since dark matter particles are collisionless, they are described in six-dimensional phase space. Because they have very small velocity dispersion, CDM particles lie on a three-dimensional hypersurface embedded in that space. One inevitable consequence of this is the formation of \textit{caustics} \cite{Zeldovich70, *Arnold82, *Shandarin89} \cite{Sikivie99} \cite{Natarajan05, *Natarajan06}. Caustics are surfaces in physical space where the density is infinite in the limit of zero velocity dispersion. In a galactic halo, both outer and inner caustics are formed. As outer caustics appear in the outer regions of the galactic halo (e.g. at a distance of order $100 \; \text{kpc}$ from the Milky Way center), they hardly affect the stellar dynamics in the disk. In this paper, we only discuss the inner caustics which form when the particles are at their closest approach to the galactic center (see Fig.~\ref{tricusp_flows}). The $n$th inner caustic forms in the flow of particles experiencing their $n$th infall in the galactic potential well. If the total angular momentum of dark matter particles is dominated by net overall rotation, each inner caustic is a closed tube whose cross section has the shape of a tricusp \cite{Sikivie98, Sikivie99} (see Fig.~\ref{tricusp}). This structure is called a \textit{caustic ring} of dark matter. 

The caustic ring model \cite{Sikivie99, Natarajan06, Duffy08} is a proposal for the full phase space distribution of cold dark matter halos. It is axially symmetric, reflection symmetric and has self-similar time evolution. It predicts that caustic rings lie in the galactic plane and that their radii $a_n(t)$ increase on cosmological time scale as $a_n(t) \propto t^{4/3}$. There is observational evidence in support of the caustic ring model \cite{Sikivie03, Banik17}. Furthermore, it was shown that net overall rotation, self-similarity and axial symmetry are the expected outcomes of the rethermalization of Bose-Einstein condensed axion dark matter \cite{Sikivie09, Erken12} before it falls onto a galactic halo \cite{Sikivie11, Banik13}. In the Milky Way, the present radius of the $n$th caustic ring is approximately $\frac{40 \; \text{kpc}}{n}$. Since the Solar System is about $8.5$ kpc away from the galactic center \cite{Binney}, the $n$ = 1, 2, 3 and 4 caustic rings have passed the solar orbit while the $n = 5$ caustic ring is approaching.

We find that star orbits are strongly perturbed by a passing caustic ring (see Fig.~\ref{Eplot}). A star on a nearly circular orbit is first attracted toward the caustic ring, then moves with and oscillates about the caustic ring for approximately $1$ Gyr, and finally returns to its original orbit (see Fig.~\ref{rtplot}). This implies that a star overdensity forms near a caustic ring. The overdensity is determined by the depth $(\Phi_c)$ of gravitational potential well of the caustic and the velocity dispersion $(\sigma)$ of the stellar population near the caustic. If $\sigma^2$ is smaller than $\Phi_c$, the stellar distribution is heavily affected by the caustic. Due to larger infall rates \cite{Duffy08}, caustic rings with smaller $n$, i.e. with larger radii, have stronger gravitational fields. Also, the stars near caustic rings of small $n$, i.e. those in the outer regions of the galactic disk, have relatively small velocity dispersions. As a result, large star overdensities form near the caustic rings of small $n$. We predict such overdensities around various caustic rings (see Figs.~\ref{overdensityX} and \ref{overdensityZ}) by simulating the dynamics of half a million stars for each. We perform the simulations for the radial and vertical motions independently as it is computationally expensive to do so for the coupled two-dimensional motions of a large number of stars.  We estimate the total overdensity near a caustic ring as the sum of overdensities formed due to the radial and vertical motions. Large star overdensities would attract more stars and interstellar gas, and are expected to be enhanced further. Such feedbacks are not considered here.

The Monoceros Ring \cite{Newberg02, *Yanny03, *Ibata03, *Rocha03} has been observed at the location of the second caustic ring. We find that the star overdensity near the $n=2$ caustic ring is of order $120\%$. This reinforces the claim of Ref.~\cite{Natarajan07} that the Monoceros Ring may be caused by the second caustic ring in our galaxy. We estimate the size of the tricusp of the $n=2$ ring to be $p \sim 2.5$ kpc based on the size of the Monoceros Ring. We find the star overdensity of order $45\%$ near the $n=3$ caustic ring, which may explain the existence of the Binney and Dehnen ring \cite{Binney97} at $13.6$ kpc. The overdensities for caustic rings with larger $n$ are smaller, e.g. approximately $30\%$ and $15 \%$ for $n=4$ and $5$ caustic rings, respectively. Such overdensities may be observed in upcoming astronomical data. Recently, three independent groups \cite{Widrow12, *Carlin13, *Williams13} have observed position dependent bulk velocities of order $10$ km/s for the stars in the extended solar neighborhood. Our work was originally motivated to investigate if the passing of a caustic ring through the solar neighborhood may explain such observations. A caustic ring passing the solar neighborhood moves with speed $\sim 1$ km/s $= 1.02$ kpc/Gyr. In Sec. \ref{IIIBa}, we show that the resultant bulk velocities of the stars are of order $\big( \frac{\Delta d}{d} \big)$ km/s, where $d$ is density of stars and $\frac{\Delta d}{d}$ their relative overdensity near the caustic ring. Even if the overdensity were of order $100 \%$, the bulk velocities are quite small compared to the observed ones. Hence, a passing caustic ring cannot explain the observed bulk velocities. More prominent astrophysical signatures of the caustic rings with large $n$ may be found in the distribution of interstellar gas. The interstellar gas has much smaller velocity dispersion than the stars and is strongly affected by caustic rings. We study the effects of the $n=5$ caustic ring on interstellar gas assuming the gas to be in thermal equilibrium in the gravitational potential of the caustic ring and of the gas itself. The caustic ring is taken to be static here because the dynamics of gas is fast compared to the time scale over which the radius of the caustic ring changes. We find that the density of the gas in a cross-sectional plane of the caustic ring has a triangular shape reminiscent of the tricusp. Triangular features in both tangent directions to the nearest caustic ring ($n=5$) have been observed in the IRAS  \cite{Sikivie03, Banik17} and GAIA\cite{Chakrabarty18} maps of the galactic plane. Interestingly, the observed features are sharper than those obtained under the above stated assumptions.

A brief outline of the paper is as follows. In Sec. \ref{II}, we describe the caustic ring model and determine the gravitational field and potential of a caustic ring. In Sec. \ref{III}, we study the dynamics of the stars in the vicinity of a caustic ring. In sec. \ref{IV}, we study its effect on the distribution of interstellar gas. Sec. \ref{V} provides a summary.

\section{\label{II}Caustic rings}

\subsection{\label{IIA}Cold dark matter caustics}

In the limit of zero velocity dispersion, dark matter particles lie on a three-dimensional hypersurface in six-dimensional phase space. As the particles in a galactic halo are huge in number, they can be labeled by a set of three continuous parameters, $\vec{\alpha} = (\alpha_1, \alpha_2, \alpha_3)$. Let $\vec{x}(\vec{\alpha},t)$ be the position of particle $\vec{\alpha}$ at time $t$. For an arbitrary physical point $\vec{r}$, let the equation $\vec{x}(\vec{\alpha},t) = \vec{r}$ have solutions $\vec{\alpha}_j (\vec{r},t)$ with $j = 1, 2, ..., N_f(\vec{r},t)$. $N_f(\vec{r},t)$ is the number of flows through $\vec{r}$ at time $t$.

If $\frac{d^3 N}{d \alpha^3} (\vec{\alpha})$ is the number density of particles in the chosen parameter space, the mass density in physical space is given by \cite{Sikivie99}
\begin{equation}
d(\vec{r},t) = m \sum_{j = 1}^{N_f(\vec{r},t)} \frac{d^3 N}{d \alpha^3} (\vec{\alpha}) \; \frac{1}{|D(\vec{\alpha},t)|}\Bigg|_{\vec{\alpha} = \vec{\alpha}_j (\vec{r},t)} \label{drt}
\end{equation}
where $m$ is the mass of each particle and $|D(\vec{\alpha},t)| \equiv |\text{det} \Big( \frac{\partial \vec{x}(\vec{\alpha},t)}{\partial \vec{\alpha}} \Big)|$ is the Jacobian of the map from $\vec{\alpha}$ to $\vec{x}$. Caustics are locations in physical space where the density $d(\vec{r},t)$ diverges because the map is singular, i.e. $D(\vec{\alpha},t) = 0$.

\subsection{\label{IIB}Flows near a caustic ring}

Each particle in an axially symmetric flow of cold dark matter is labeled by two parameters ($\alpha, \tau$). The third parameter labeling a particle is its azimuth which is irrelevant in the case of axial symmetry. $\alpha$ is the angle from the $z=0$ plane at the time of the particle's most recent turnaround, i.e. $\alpha = \frac{\pi}{2} - \theta$ where $\theta$ is the polar angle in spherical coordinates. For each value of $\alpha$, $\tau = 0$ is defined as the time when the particle crosses the $z=0$ plane, i.e. $z(\alpha, \tau = 0) = 0$. The coordinates $(\rho, z)$ of a particle near the caustic ring are given by \cite{Sikivie99}
\begin{eqnarray}
\rho && = a + \frac{1}{2} u (\tau - \tau_{0})^{2} - \frac{1}{2} s \alpha^{2} \label{rho} \\
z && = b \alpha \tau \label{z} 
\end{eqnarray}
where $a, \; u, \; \tau_{0}, \; s$ and $b$ are constants for a given caustic. The caustic occurs where the Jacobian $|D_{2} (\alpha, \tau)| \equiv |\text{det} \frac{\partial(\rho, z)}{\partial(\alpha, \tau)}|$ is $0$. Its location in the $\rho$-$z$ plane as a function of the parameter $\tau$ is given by
\begin{eqnarray}
\rho &&= a + \frac{1}{2} u (\tau - \tau_{0})(2\tau - \tau_{0}) \label{rho_tau}  \\
z &&= \pm b \sqrt{\frac{u}{s} \tau^{3} (\tau_{0} - \tau)} \;\; .\label{z_tau}
\end{eqnarray}
Equations (\ref{rho_tau}) and (\ref{z_tau}) define the tricusp (see Figs.~\ref{tricusp_flows} and \ref{tricusp}). Its size is $p$ in the $\rho$-direction and $q$ in the $z$-direction: $p = \frac{1}{2} u \tau_{0}^{2}, \;\;\; q = \frac{\sqrt{27}}{4} \frac{b}{\sqrt{us}} p$. The density in physical space is \cite{Sikivie99}
\begin{equation}
d(\rho, z) = \frac{1}{\rho} \sum_{j = 1}^{N_f(\vec{r},t)} \frac{dM}{d\Omega d\tau} (\alpha, \tau) \frac{cos \alpha}{|D_{2} (\alpha, \tau)|} \Bigg|_{\alpha_j (\rho, z), \tau_j (\rho, z)} \; , \label{drhoz}
\end{equation}
where $\frac{dM}{d\Omega d\tau}$ is the mass of dark matter particles falling in per unit solid angle per unit time.

\subsection{\label{IIC}Self-similarity}

The evolution of dark matter particles in a galactic halo is self-similar if there is no special time in its history \cite{Gunn77}\cite{Fillmore84}\cite{Bert85}. Self-similarity implies  that the phase space distribution remains identical to itself except for an overall rescaling of its density, size and velocity. As pointed out in Ref.~\cite{Duffy08}, self-similarity does not require any symmetry. In the self-similar model, the radius of each caustic ring increases on cosmological time sale as $a(t) \propto t^{\beta}$ where $\beta = \frac{2}{3} + \frac{2}{9\epsilon}$. From the slope of the power spectrum of density perturbation on galactic scales, $\epsilon$ is determined to be in the range $0.25 < \epsilon < 0.35$ \cite{Sikivie97}. The value $\epsilon = \frac{1}{3}$ (hence $\beta = \frac{4}{3}$) is consistent with the observational evidence for caustic rings, the slope of the power spectrum on galactic scales and with the value from tidal torquing \cite{Sikivie11}.

\subsection{\label{IID}Gravitational field of a caustic ring}

The gravitational field inside and near a caustic ring has been calculated in Refs.~\cite{Sikivie99, Tam12}. Assuming the size ($p, q$) of the cross section of the ring to be much smaller than the radius $a$ of the ring, the gravitational field in terms of the rescaled variables $\big( X = \frac{\rho - a}{p}, Z = \frac{z}{p}\big)$ is given by
\begin{equation}
\vec{g} = -8\pi G\; \frac{dM}{d\Omega d\tau} \; \frac{1}{b \rho} [I_{\rho} (X, Z) \hat{\rho} + I_{z} (X, Z) \hat{z}] \; , \label{gfield3}
\end{equation}
where 
\begin{eqnarray}
I_{\rho} &&= \frac{1}{2\pi} \int_{-\infty}^{\infty} dA \int_{-\infty}^{\infty} dT \; \frac{X - (T - 1)^{2} + \xi A^{2}}{(X - (T - 1)^{2} + \xi A^{2})^{2} + (Z - 2AT)^{2}} \;\; ,\label{I_rho} \nonumber \\
I_{z} &&= \frac{1}{2\pi} \int_{-\infty}^{\infty} dA \int_{-\infty}^{\infty} dT \; \frac{(Z - 2AT)}{(X - (T - 1)^{2} + \xi A^{2})^{2} + (Z - 2AT)^{2}} \;\; \label{I_z} \nonumber
\end{eqnarray}
and $\xi = \frac{su}{b^{2}}$. The variables $(\alpha, \tau)$ have been replaced by $\big( A = \frac{b \alpha}{u \tau_{0}}, T = \frac{\tau}{\tau_{0}} \big)$. In Ref.~\cite{Tam12}, $\xi$ was taken to be unity and the integrals were calculated both analytically, using residue theory, and numerically with consistent results. Equation (\ref{gfield3}) gives the gravitational field of the whole flow forming the caustic. The gravitational field due to the flows without caustics is that of a smooth halo which results in a flat rotation curve. We are interested in the modified gravitational field only due to the formation of the caustic. In the caustic ring model, one cannot remove a flow without removing the caustic in the flow. To separate the gravitational field of a caustic ring from that of the flow of which it is part, we introduce a long distance damping factor,
\begin{equation}
\vec{g}_{c} (\rho, z) = \exp \Big(- \frac{s^2}{R^2}\Big) \; \vec{g} (\rho, z) \label{g_c}
\end{equation}
where $s^2 = (\rho - a - \frac{p}{4})^2 + z^2$ is the distance of the point $(\rho, z)$ from the center of the caustic and $R \sim 1.5 p$ is the distance scale over which the effects of the caustic ring are significant. 

\begin{figure}
\begin{center}
\subfigure[]{
\includegraphics[scale=0.5]{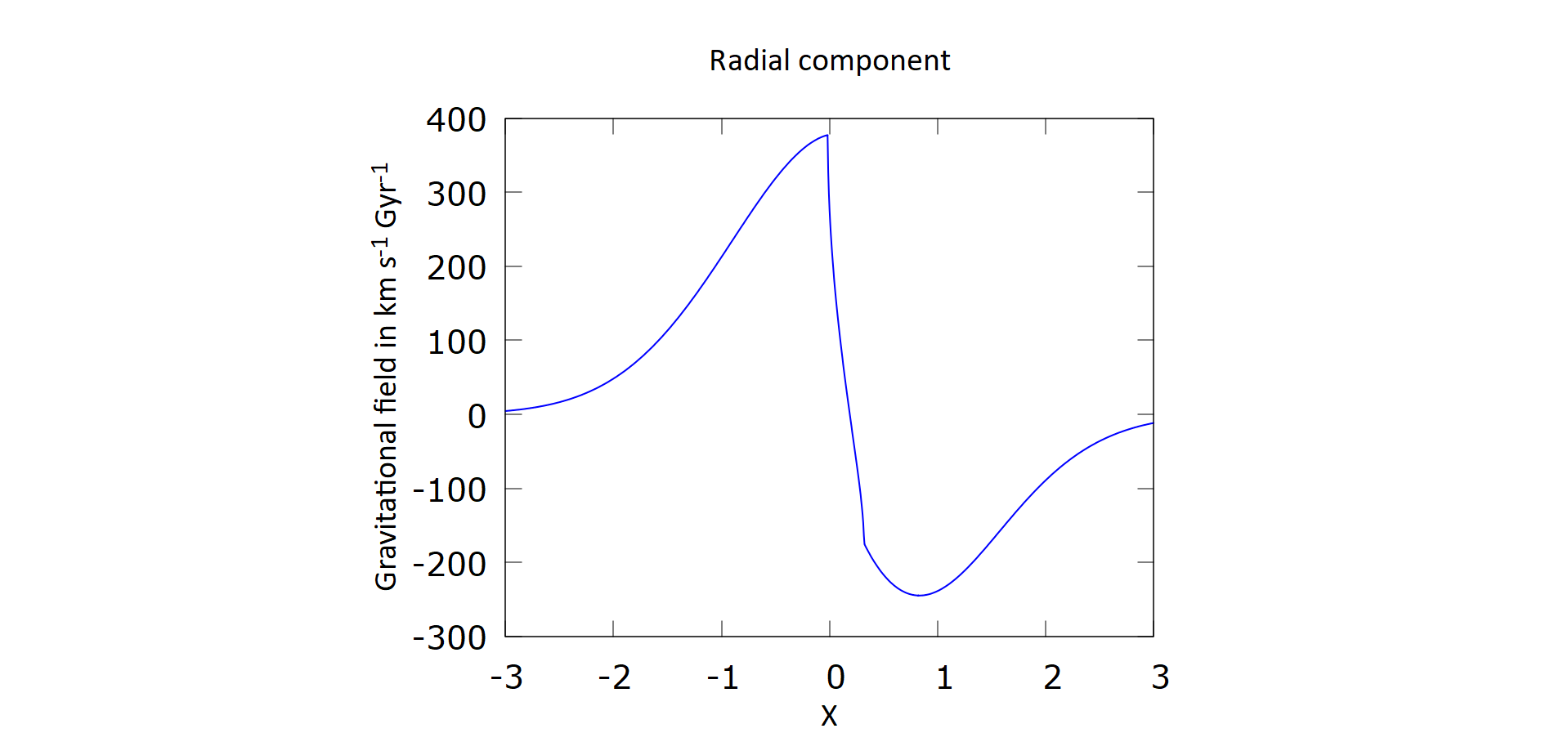}
}
\subfigure[]{
\includegraphics[scale=0.5]{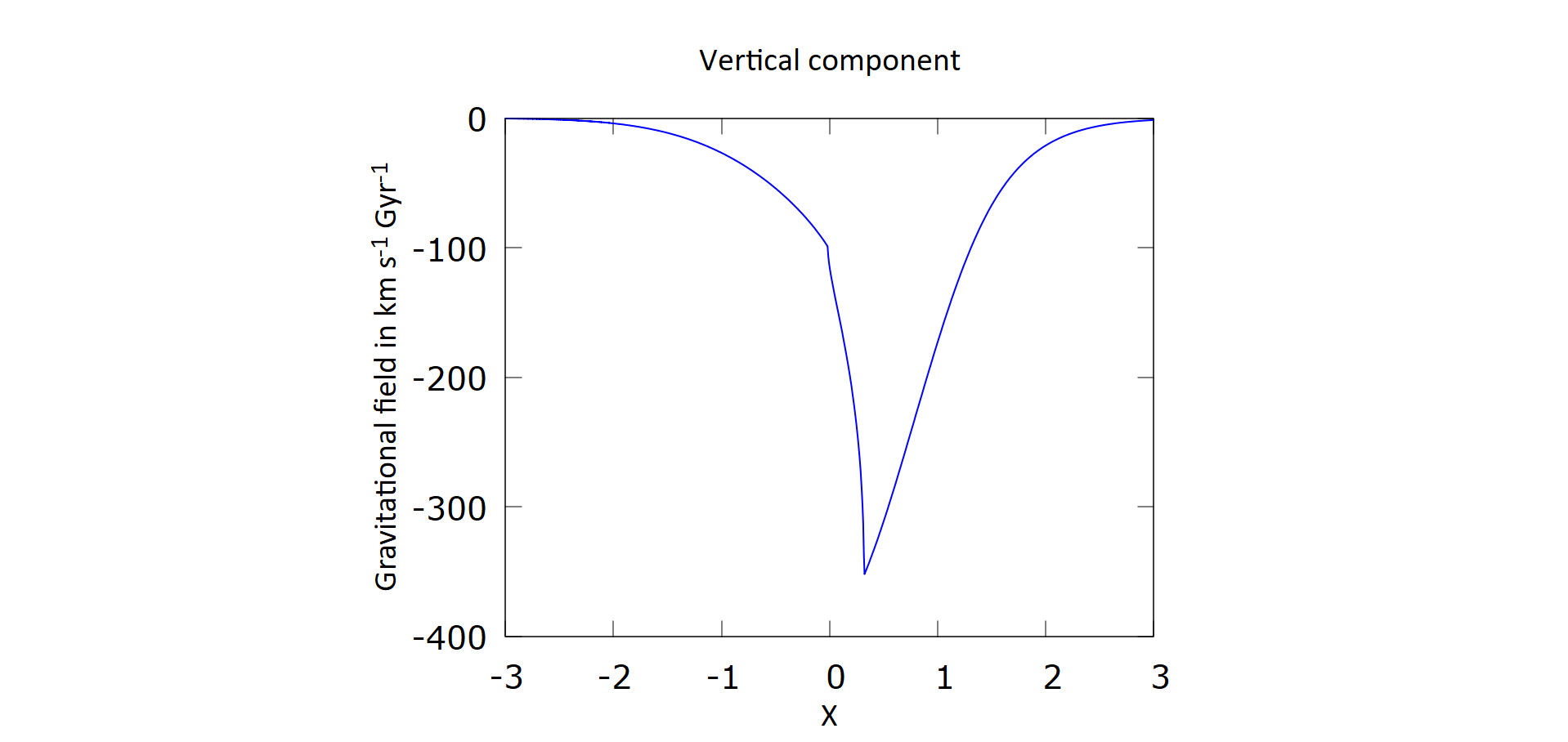}
}
\end{center}
\caption{\label{gplot}The radial and vertical components of gravitational field $\vec{g}_{c} (X, Z = 0.25)$ [see Eq. (\ref{g_c})] due to a caustic ring as a function of $X = \frac{\rho - a}{p}$  for $Z = \frac{z}{p} = 0.25$. The parameters of the caustic ring have been chosen as $a = 8.0$ kpc, $p = 0.5$ kpc, $b = 523$ km/s, $\frac{dM}{d\Omega d\tau} = 32 \frac{M_{\odot}}{\text{sterad. yr}}$. The chosen parameters are similar to that of the $n=5$ caustic ring, the one closest to us.}
\end{figure}

The parameters for various caustic rings have been listed in Ref.~\cite{Duffy08}. The values for the infall rate $\frac{dM}{d\Omega d\tau}$ given in Ref.~\cite{Duffy08} are based on the assumption of isotropic infall of dark matter particles. However, Ref.~\cite{Banik13} argues that, because the vortices in the axion Bose-Einstein condensate attract each other, numerous smaller vortices join to form a huge vortex along the rotation axis of the galaxy. As a result, axions fall in preferentially along the galactic plane and caustic rings are enhanced. This explains why the bumps in the Milky Way rotation curve at the locations of the caustic rings are typically a factor of $4$ larger than that attributed only to the caustic rings with isotropic infall \cite{Sikivie03}. We therefore multiply the infall rates in Ref.~\cite{Duffy08} by a factor $4$ to account for the formation of the `big vortex', giving them the values
\begin{equation}
\Big(\frac{dM}{d\Omega d\tau}\Big|_n : n = 1, 2, 3, 4, 5, ... \Big) \approx ( 210, 95, 60, 40, 32, ...)\; \frac{M_{\odot}}{\text{sterad-yr}}. \label{infallratecaustics}
\end{equation}
In this paper, unless mentioned otherwise, we use the following parameters to describe a caustic ring passing through the solar neighborhood: $p = 0.5$ kpc, $b = 523$ km/s, $V = \frac{da}{dt} = 1$ kpc/Gyr and $\frac{dM}{d\Omega d\tau} = 32 \frac{M_{\odot}}{\text{sterad-yr}}$. The chosen infall rate is similar to that of the $n=5$ caustic ring, which is the one closest to us. In Fig.~\ref{gplot}, we show the radial and vertical components ($g_{c\rho}$, $g_{cz}$) of the gravitational field at $Z = \frac{z}{p} = 0.25$ as a function of $X = \frac{\rho - a}{p}$.

\subsection{\label{IIE}Gravitational potential of a caustic}

Assuming the size of the cross section of a caustic ring to be much smaller than its radius (i.e. $p, q \ll a$; see Fig.~\ref{tricusp}), the gravitational potential $\Phi(X, Z)$ near the caustic ring is given by
\begin{equation}
\Phi(X, Z) - \Phi(X_0, Z_0) = 2 G \frac{dM}{d\Omega d\tau} \frac{p}{ba} J(X, Z; X_0, Z_0) \label{phixz}
\end{equation}
where
\begin{equation}
J(X, Z; X_0, Z_0) = \int_{-\infty}^{\infty} dA \int_{-\infty}^{\infty} dT \ln \frac{(X - (T-1)^2 + A^2)^2 + (Z-2AT)^2}{(X_0 - (T-1)^2 + A^2)^2 + (Z_0-2AT)^2} \;\; . \label{jxz}
\end{equation}
We choose the center of the tricusp as the reference point ($X_0, Z_0$). Figure \ref{causticpot} shows the two-dimensional plot of the caustic potential
\begin{equation}
\Phi_c (X, Z) = \Phi(X, Z) - \Phi(X_0 = 0.25, Z_0 = 0) . \label{phicxz}
\end{equation}
It looks smooth and continuous, although its second derivative $\nabla^2 \Phi_c$ diverges at the caustic.

\section{\label{III}Effects on stars}

\subsection{\label{IIIA}A single star}

For the stars near the galactic disk, small radial and vertical oscillations can be treated independently \cite{Binney} in the absence of a caustic. The effective potential for radial motion consists of a logarithmic gravitational potential and the angular momentum barrier:
\begin{equation}
\Phi_{\text{eff}} (\rho) = v_{\text{rot}}^2 \ln \rho + \frac{l^2}{2 \rho^2} \label{Phi_eff}
\end{equation}
where $l = \rho^2 \dot{\phi} = \text{constant}$. For small radial oscillations about the minimum $\rho_0 = \frac{l}{v_{\text{rot}}}$ of the effective potential, the angular frequency is $\omega_{\rho} = \frac{\sqrt{2}v_{\text{rot}}}{\rho_0}$. For vertical motion, we choose the potential $\Phi_z (z)$ of an isothermal stellar disk with velocity dispersion $\sigma_{z}$ and scale height $z_0$ \cite{Binney}
\begin{equation}
\Phi_z (z) = \sigma_{z}^2 \; \ln \Big[\cosh^2 \Big( \frac{z}{2z_0} \Big) \Big] . \label{Phi_z}
\end{equation}
In the presence of a passing caustic ring, the radial and vertical accelerations of a star are given by
\begin{eqnarray}
a_{\rho} (\rho, z) &&= - \frac{v_{\text{rot}}^2}{\rho} + \frac{l^2}{\rho^3} + g_{c\rho} (\rho, z, a(t)) \\ \label{arho_c}
a_z (\rho, z) &&= - \frac{\sigma_{z}^2}{z_0} \tanh \Big( \frac{z}{2z_0} \Big) + g_{cz}(\rho, z, a(t)) \label{az_c}
\end{eqnarray}
where $g_{c\rho} (\rho, z, a(t))$ and $g_{cz}(\rho, z, a(t))$ are the contributions [see Eqs. (\ref{gfield3}) and (\ref{g_c})] from the caustic ring with radius $a(t)$. To visualize the effects of the caustic ring on the star, we define radial and vertical energies per unit mass
\begin{eqnarray}
E_{\rho} &&= \frac{1}{2} v_{\rho}^2 + \Phi_{\text{eff}} (\rho) - \Phi_{\text{eff}} (\rho_0) \;\; , \label{E_rho} \\
E_z &&=  \frac{1}{2} v_z^2 + \Phi_{z} (z) \;\; , \label{E_z}
\end{eqnarray}
where $\Phi_{\text{eff}} (\rho)$ and $\Phi_{z} (z)$ are given by Eqs. (\ref{Phi_eff}) and (\ref{Phi_z}) and $\rho_0$ is the minimum of $\Phi_{\text{eff}}(\rho)$. $E_{\rho}$ and $E_z$ remain constant in the absence of a caustic because radial and vertical motions are independent. 

We numerically solve the equations of motion of a star in the $\rho$-$z$ plane with arbitrary initial conditions as the caustic ring passes through its orbit. The velocity dispersions of the stars in the solar neighborhood are $40$ km/s in the radial direction and $20$ km/s in the vertical direction, and the scale height $z_0 = 0.5$ kpc \cite{Binney}. As an example, we choose a star that orbits the galaxy at $\rho_0 = 8$ kpc with small radial and vertical oscillations such that $v_{\rho}^{\text{max}} = 10$ km/s and $v_z^{\text{max}} = 5$ km/s. In Fig.~\ref{Eplot}, we show the energies, $E_{\rho}$ and $E_z$, of the star as the caustic ring passes through its orbit. The fluctuations in the energies are large and occur on a time scale of approximately $2$ Gyr. The fluctuations are smaller for larger initial values of $E_{\rho}$ and $E_z$.

The stars which are most affected by a passing caustic ring are those with nearly circular orbits. In Fig.~\ref{rtplot}, we plot the radial coordinate $\rho$ of a star, and the locations of the rear ($a$) and front ($a+p$) of the tricusp of the caustic ring as a function of time. Initially, the star in Fig.~\ref{rtplot}(a) has an exactly circular orbit with radius $\rho_0 = 8$ kpc, whereas the star in Fig.~\ref{rtplot}(b) has an almost circular orbit with $\rho_0 = 8$ kpc and $v_{\rho}^{\text{max}} = v_z^{\text{max}} = 5$ km/s. The tricusp moves radially outward with a speed $\frac{da}{dt} = 1$ kpc/Gyr with $a(t=0) = 6$ kpc. In both cases, as the figures show, the star is first attracted towards the tricusp and then moves with and oscillates about the tricusp for approximately $1$ Gyr before returning to its initial orbit due to conservation of the angular momentum. For the chosen caustic ring parameters, we find that all the stars with $v_{\rho}^{\text{max}}, v_z^{\text{max}} \leq 10$ km/s exhibit such behavior. The intermediate phase of following the tricusp causes stellar overdensities around the caustic ring. 

\subsection{\label{IIIB}Distribution of stars}

When a caustic ring passes through a relaxed distribution of stars, it generates bulk velocities of the stars and perturbs the density profile of the stellar population.

\subsubsection{\label{IIIBa} Bulk velocities}

From the continuity equation, $\frac{\partial d}{\partial t} + \vec{\nabla}. (d\; \vec{v}) = 0$, the bulk velocities of the stars are of order,
\begin{equation}
v \sim \frac{\Delta d}{d} \; \frac{\Delta x}{\Delta t} \;\; , \label{v_bulk}
\end{equation}
where $\frac{\Delta d}{d}$ is the relative overdensity caused by the caustic and, $\Delta x$ and $\Delta t$ are the length and time scales over which the stellar distribution changes. For a caustic ring with radius $a$ and tricusp of size $p$, $\Delta x \sim p$ and $\Delta t \sim \frac{p}{V}$ where $V \sim \frac{da}{dt}$ is the speed of the tricusp in the radial direction. All the caustic rings move slowly with speed $V \sim 1$ km/s. Hence, the bulk velocities are $v \sim 1\; \text{km/s} \; \big(\frac{\Delta d}{d}\big)$. Even if the overdensities are as large as $\sim 100$\%, the bulk velocities induced by the caustic rings cannot be more than a few km/s. This is too small to explain the recently observed \cite{Widrow12, *Carlin13, *Williams13} bulk velocities of order $10$ km/s for the stars in the solar neighborhood.

\subsubsection{\label{IIIBb} Overdensities}

A stellar population with smaller velocity dispersion is more susceptible to a passing caustic ring. The present locations of the caustic rings are given by
\begin{equation}
(a_n : n = 1, 2, 3, 4, 5, ...) \approx (40, 20, 13.3, 10, 8, ...) \; \text{kpc}. \label{ancaustics}
\end{equation}
The radial velocity dispersion $\sigma_{\rho}$ of the stars at a distance $\rho$ from the galactic center decays exponentially with $\rho$ \cite{Lewis89, *Kubryk15}: 
\begin{equation}
\sigma_{\rho} (\rho) \approx (40\; \text{km/s})\;\text{exp}\Big[- \frac{(\rho - 8.5 \; \text{kpc})}{8 \; \text{kpc}}\Big]. \label{sigmavar}
\end{equation}
Therefore, the radial velocity dispersions of the stars near the first five caustic rings are: 
\begin{equation}
(\sigma_{n\rho} : n = 1, 2, 3, 4, 5, ...) \approx (1, 10, 20, 30, 40, ...)\; \text{km/s}. \label{sigmacaustics} 
\end{equation}
The vertical velocity dispersions $\sigma_{z}$ are typically half of the radial ones. The infall rates for the caustic rings are given by Eq. (\ref{infallratecaustics}). Caustic rings with smaller $n$ have stronger gravitational field [see Eq. (\ref{gfield3})] and are surrounded by stars with smaller velocity dispersions.

In the presence of a caustic ring, the stellar dynamics in the radial and vertical directions are not independent. Simulating the dynamics of a large number of stars in the $\rho$-$z$ plane near the caustic ring is computationally expensive. For each caustic ring, we simulate one-dimensional motion of the stars in the radial and vertical directions independently. To determine how the star overdensities change with radial (vertical) coordinates, we simulate the radial (vertical) motions and suppress the vertical (radial) dynamics of the stars. The initial motion in the radial direction is assumed to be simple harmonic, i.e. a star oscillating about $\rho = \rho_0$ moves in a harmonic potential with $\omega_{\rho} = \frac{\sqrt{2} v_{\text{rot}}}{\rho_0}$ where $v_{\text{rot}} = 220$ km/s [see Eq. (\ref{Phi_eff})]. For the radial motion of the stars near each caustic ring, we generate a relaxed distribution of $500,000$ stars with phase space density $f(v_{\rho}, \rho) \sim \text{exp}\Big(- \frac{v_{\rho}^2 + \omega_{\rho}^2 (\rho - \rho_0)^2}{2\sigma_{\rho}^2}\Big)$. The initial vertical motion is determined by the potential $\Phi_z(z)$ in Eq. (\ref{Phi_z}) and the phase space density is given by $f(v_z, z) \sim \text{sech}^2 \Big( \frac{z}{2z_0} \Big) \text{exp}\Big(- \frac{v_{z}^2}{2\sigma_{z}^2}\Big)$. In the absence of the caustics, each stellar distribution remains in equilibrium; i.e. the number density profile $d_{\text{eq}}(\rho)$ or $d_{\text{eq}}(z)$ does not change with time. 

For the $n$th caustic ring, the parameters for the radial and vertical dynamics of the stellar distributions are chosen as follows: $\sigma_{\rho} = 2\sigma_{z} = \sigma_{n\rho}$ [see Eq. (\ref{sigmacaustics})], $\langle \rho \rangle = \rho_0 = a_n$ [see Eq. (\ref{ancaustics})] and $z_0 = 0.5$ kpc. To minimize the error due to finite size of the stellar population, we choose the size $p$ of the tricusp of each caustic ring to be much smaller than the size of the corresponding stellar population. While simulating the dynamics of the stars as the caustic rings pass through them, we take several snapshots of each stellar distribution and determine the relative overdensity $\frac{\Delta d}{d} = \frac{d - d_{\text{eq}}}{d_{\text{eq}}}$. We find that, as long as the size $p$ of the tricusp is much smaller than the spread of the stellar population, the overdensity is independent of the size $p$. In Fig.~\ref{overdensityX}, we plot the stellar overdensities $\frac{\Delta d}{d}$ due to the radial motion at the locations of different caustic rings as functions of the rescaled radial coordinate $X = \frac{\rho - a}{p}$. In Fig.~\ref{overdensityZ}, we plot the same for the vertical motion as functions of the rescaled vertical coordinate $Z = \frac{z}{p}$. We did not consider the $n = 1$ caustic ring at $40$ kpc. In the linear approximation, the total overdensity near the caustic is the sum of the overdensities obtained for both directions. We find the maximum total overdensity to be of order $120\%, 45\%, 30\%$ and $15\%$ for the $n = 2, 3, 4$ and $5$ caustic rings. If the star overdensity near a caustic ring is large, it enhances the effects of the caustic by attracting more stars and interstellar gas. We did not include such backreactions in our simulation here. The large star overdensities near the $n = 2$ and $3$ caustic rings may explain the existence of the Monoceros Ring \cite{Newberg02, *Yanny03, *Ibata03, *Rocha03} at $20$ kpc and the Binney and Dehnen Ring \cite{Binney97} at $13.6$ kpc. The Monoceros Ring has a vertical scale height of order $10$ kpc \cite{Newberg02}. According to Fig.~\ref{overdensityZ}, the vertical scale height of the overdensity due to the $n=2$ caustic ring is of order $4p$. So, to form an overdensity of vertical size 10 kpc, the $n=2$ caustic ring is required to have a size $p \sim 2.5$ kpc. The sizes of the $n=1$ and $2$ caustic rings are not known from the bumps in the rotation curve while the size of the third caustic has been determined to be 1 kpc \cite{Duffy08}. Since the overdensities near the $n = 4$ and $5$ caustic rings occur within a distance of few kpc from the Sun, they may be observed in upcoming astronomical data such as from GAIA.

\section{\label{IV}Effects on interstellar gas}

The bulk properties of a distribution of stars in the solar neighborhood are affected by the caustic rings only at the $15\%$ level because the square of the velocity dispersion ($\sigma^2$) of the stars is larger than the depth of the potential ($\Phi_c$) of a caustic ring. Since gas and dust in the interstellar medium have smaller velocity dispersions \cite{Binney}, their bulk properties, e.g. density, are expected to be affected more. Assuming the gas to be in thermal equilibrium in the gravitational potential of both the caustic and the gas itself, we study its density profile near the $n=5$ caustic ring. The caustic is taken to be static as the gas dynamics is fast compared to the time scale over which the radius of the caustic ring changes. 

In the absence of the caustic ring, the interstellar gas is assumed to be in thermal equilibrium through self-gravitational interactions. Its potential and density are given by  \cite{Binney}:
\begin{eqnarray}
\Phi_g (\rho, z) &&= \sigma_g^2 \; \ln \Big[\cosh^2 \Big( \frac{z}{2z_g} \Big) \Big] \label{Phigas} \\
d_g (\rho, z) &&= d^{0}_{g} \; \text{sech}^2 \Big( \frac{z}{2z_g} \Big) \label{rhogas}
\end{eqnarray}
where $z_g = \frac{\sigma_g}{\sqrt{8\pi G d^{0}_{g}}}$ is the scale height and $\sigma_g$ is the velocity dispersion. The parameters in the solar neighborhood are the following: $z_g \approx 65$ pc, $\sigma_g \approx 5$ km/s and $d^{0}_{g} \approx 0.05 \; \text{M}_{\odot}/ \text{pc}^{3}$ \cite{Binney}. Since the scale height of the gas is much smaller than that of disk stars ($300$ pc for thin and $500$ pc for thick disks), we ignore the gravity of the stars. The potential and density are taken to be independent of the radial coordinate $\rho$ since they do not change significantly with $\rho$ over the size of the tricusp.

We change our coordinate system from $(\rho, z)$ to $(X, Z) = (\frac{\rho - a}{p}, \frac{z}{p})$ as the latter is more convenient. In the presence of a caustic ring, the potential $\Phi (X, Z)$ due to gas is the solution of Poisson's equation,
\begin{equation}
\nabla^2 \Phi (X, Z) = 4\pi G d_g (X, Z) \label{Poisson}
\end{equation}
where 
\begin{equation}
d_g (X, Z) = d_g (X_0, Z_0) \exp \Big(-\frac{\Phi(X, Z) + \Phi_c (X, Z)}{\sigma_g^2}\Big)  \label{rhoxz}
\end{equation}
assuming $\Phi(X_0, Z_0) = \Phi_c (X_0, Z_0) = 0$. We calculate the new potential $\Phi (X, Z)$ using a two-dimensional Green's function. In the $X$-$Z$ plane, we choose a sufficiently large region around the tricusp such that, at the boundary of the region, $\Phi (X, Z)$ tends to be $\Phi_g (X, Z)$ [see Eq. (\ref{Phigas})]. Choosing a Green's function $G(X, Z; X', Z')$ that vanishes at the boundary, we have
\begin{equation}
\Phi (X, Z) = \int dX' \int dZ' \; G(X, Z; X', Z') 4\pi G d_g (X', Z') + \oint dl' \; \Phi_g (X', Z') \frac{\partial G}{\partial n'}(X, Z; X', Z') . \label{phixzn}
\end{equation}
We solve the above equation numerically. The solution for $\Phi(X, Z)$ converges after few iterations. In Fig.~\ref{densityplot}, we plot the density of gas $d_g (X, Z)$ near the tricusp of the $n=5$ caustic ring with size $p = 150$ pc. As evident from the figure, the density profile has a triangular shape as does the caustic potential shown in Fig.~\ref{causticpot}. In the recent GAIA sky map of the Milky Way \cite{GAIADR1, *GAIADR2}, triangular features are observed \cite{Chakrabarty18} in both tangent directions to the fifth caustic ring. However, the observed features have sharper edges than the triangular shape of Fig.~\ref{densityplot}.

\section{\label{V}Summary}

\begin{figure}
\begin{center}
\subfigure[]{
\includegraphics[scale=0.5]{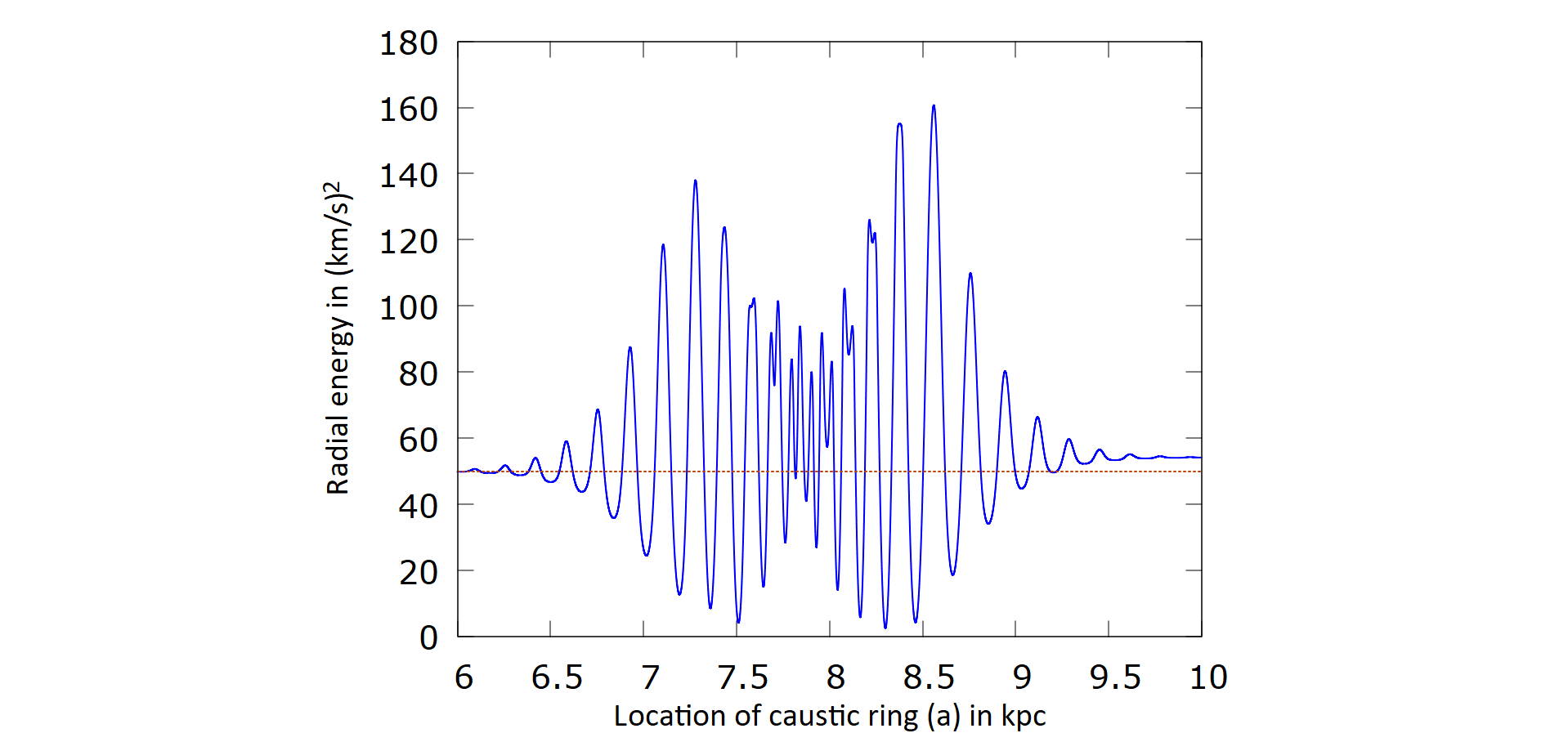}
}
\subfigure[]{
\includegraphics[scale=0.5]{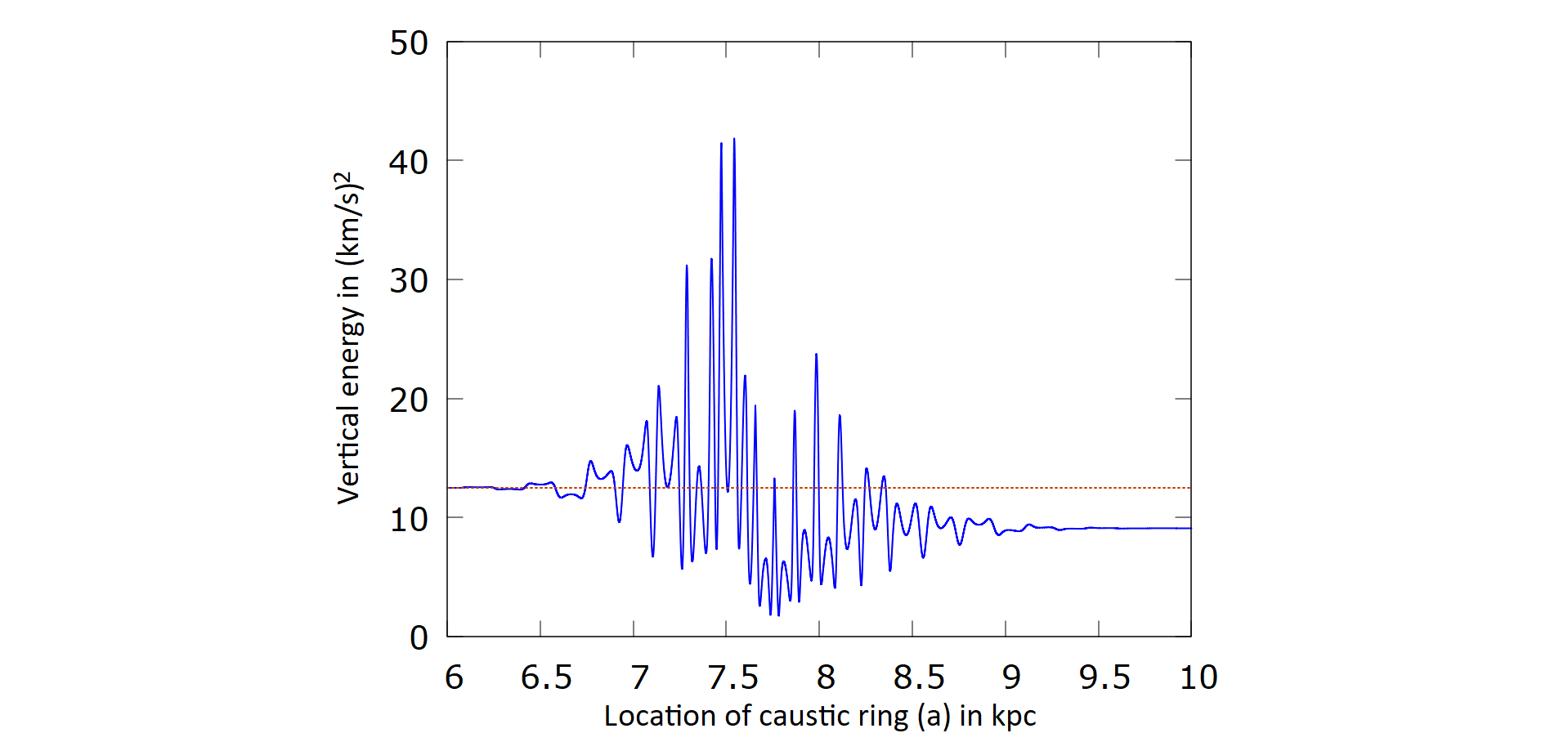}
}
\end{center}
\caption{\label{Eplot}Variation of the radial and vertical energies, $E_{\rho}$ and $E_z$ (see Eqs.~\ref{E_rho}-\ref{E_z}), of a star when a caustic ring passes through its orbit. The star was initially orbiting the galaxy at $\rho_0 = 8$ kpc oscillating in the vertical direction with $v_{z}^{\text{max}} = 5$ km/s and in the radial direction with $v_{\rho}^{\text{max}} = 10$ km/s. The dotted lines indicate the values of $E_{\rho}$ and $E_z$ in the absence of the caustic ring. Fluctuations in the energies are shown as the radius $a(t)$ of the caustic ring changes from $6$ kpc to $10$ kpc. As the caustic ring moves very slowly ($1$ kpc/Gyr), the time scale over which the fluctuations occur is very large (approximately $2$ Gyr).}

\end{figure}

\begin{figure}
\begin{center}
\subfigure[]{
\includegraphics[scale=0.5]{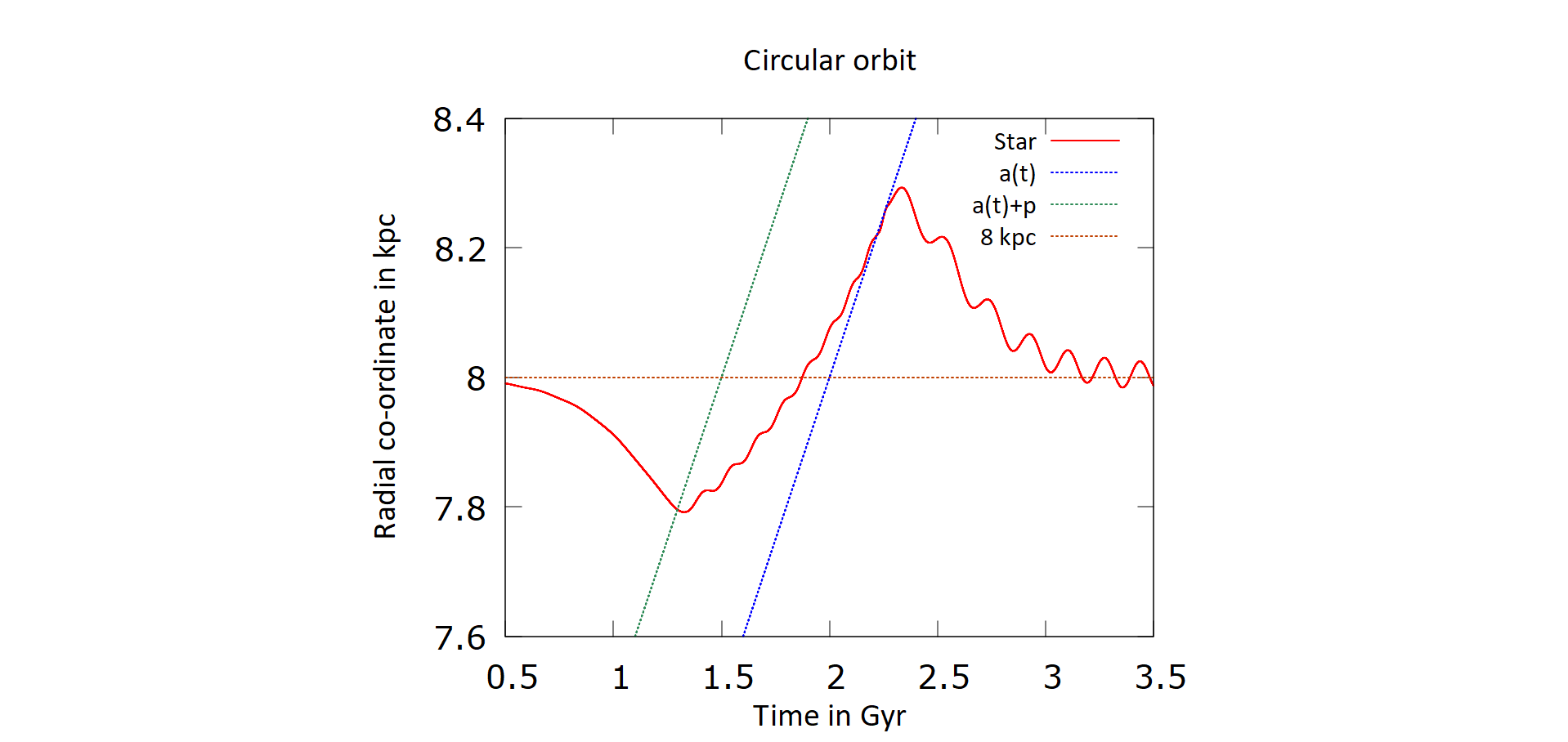}
}
\subfigure[]{
\includegraphics[scale=0.5]{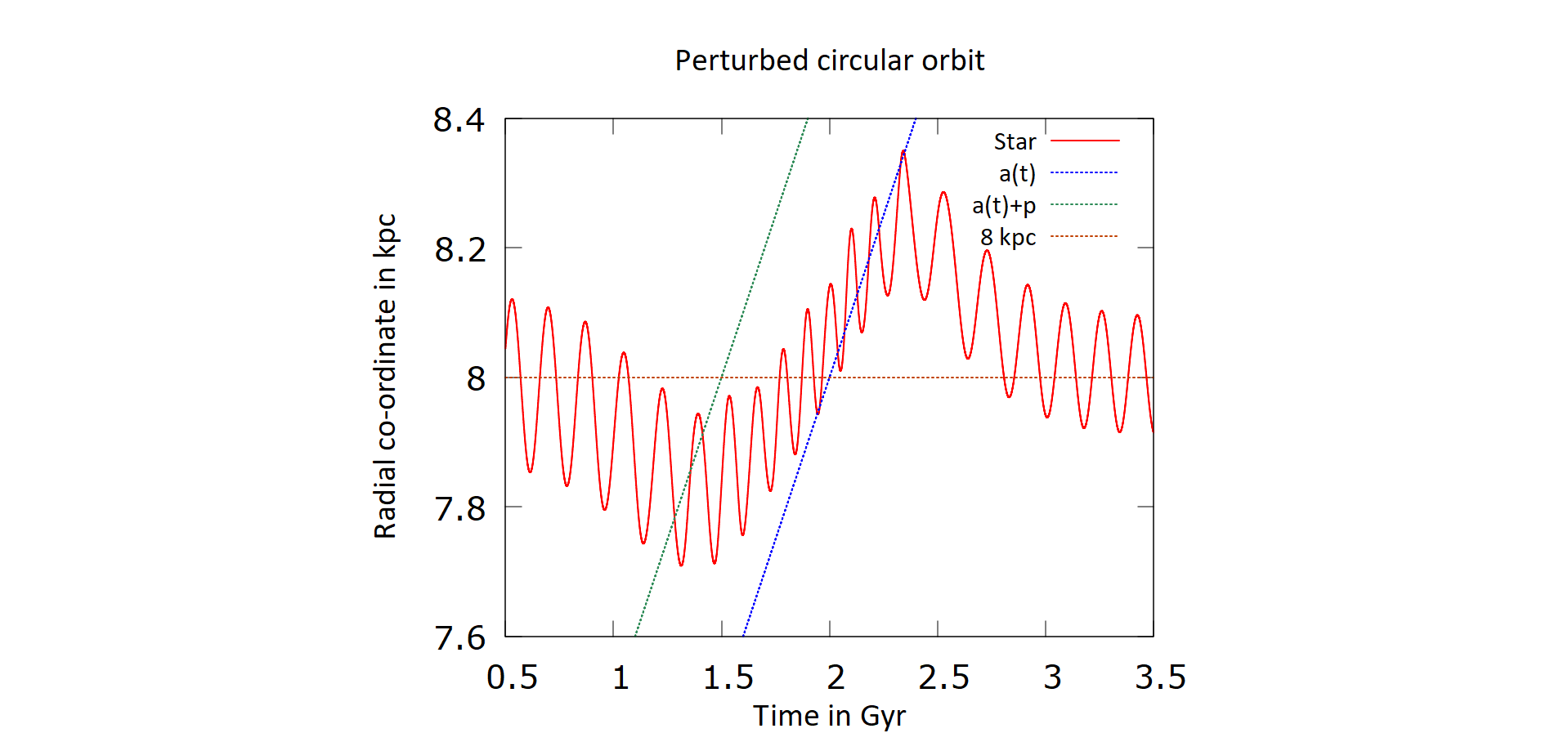}
}
\end{center}
\caption{\label{rtplot}Variation of the radial coordinate $\rho$ of a star as a caustic ring with $p = 0.5$ kpc passes through its orbit. In the absence of the caustic ring, the star in Fig.~(a) has an initially circular orbit of radius $\rho_0 = 8$ kpc whereas the star in Fig.~(b) has an initially almost circular orbit with $\rho_0 = 8$ kpc and $v_{\rho}^{\text{max}} = v_z^{\text{max}} = 5$ km/s. The thick dotted lines, $a$ and $a+p$, indicate the rear and front of the tricusp (see Fig.~\ref{tricusp}) moving with speed $1$ kpc/Gyr. In both cases, the star is initially attracted towards the tricusp, then moves with and oscillates about the tricusp for approximately $1$ Gyr before coming back to its initial orbit.}
\end{figure}

The radii $a_n$ of caustic rings increase slowly at the rate of approximately $1\; \text{km/s} \; \big(\frac{a_n}{8\; \text{kpc}}\big)$. We study the dynamics of stars and interstellar gas as the caustic rings pass through them. We find that stellar orbits are highly perturbed by a passing caustic ring (see Fig.~\ref{Eplot}). A star moving in a nearly circular orbit is first attracted towards the caustic ring, then moves with and oscillates about the caustic for approximately $1$ Gyr, and finally returns to its original orbit (Fig.~\ref{rtplot}) as a result of angular momentum conservation.

\begin{figure}
\includegraphics[scale=0.5]{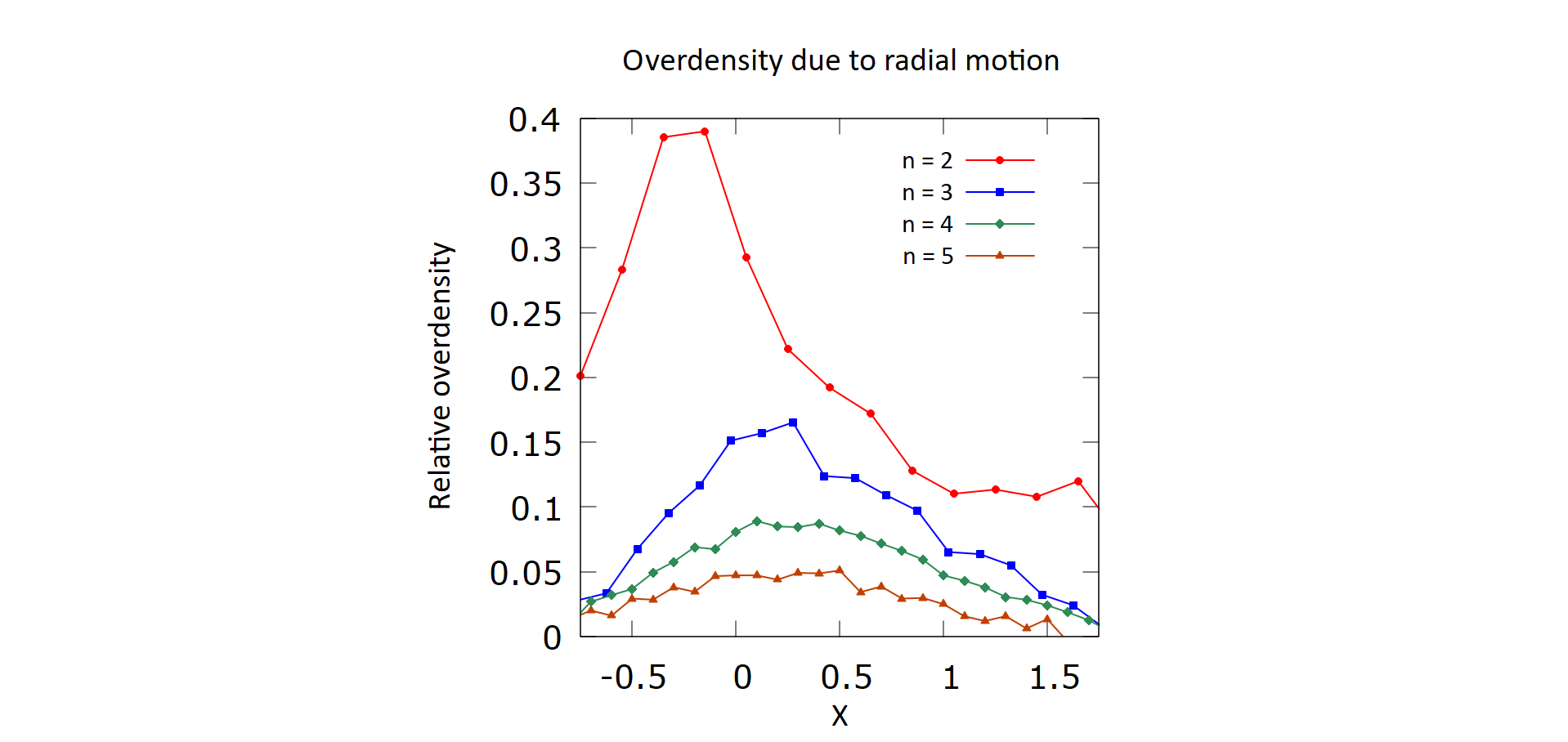}
\caption{\label{overdensityX}Relative overdensities of the stars $\frac{\Delta d}{d} = \frac{d - d_{\text{eq}}}{d_{\text{eq}}}$ due to radial motion near the $n = 2, 3, 4$ and $5$ caustic rings as functions of the rescaled radial coordinate $X = \frac{\rho - a}{p}$. Stellar overdensities are higher near the caustic rings with smaller $n$ because they have stronger gravitational field and are surrounded by stars with lower velocity dispersions.}
\end{figure}

\begin{figure}
\includegraphics[scale=0.5]{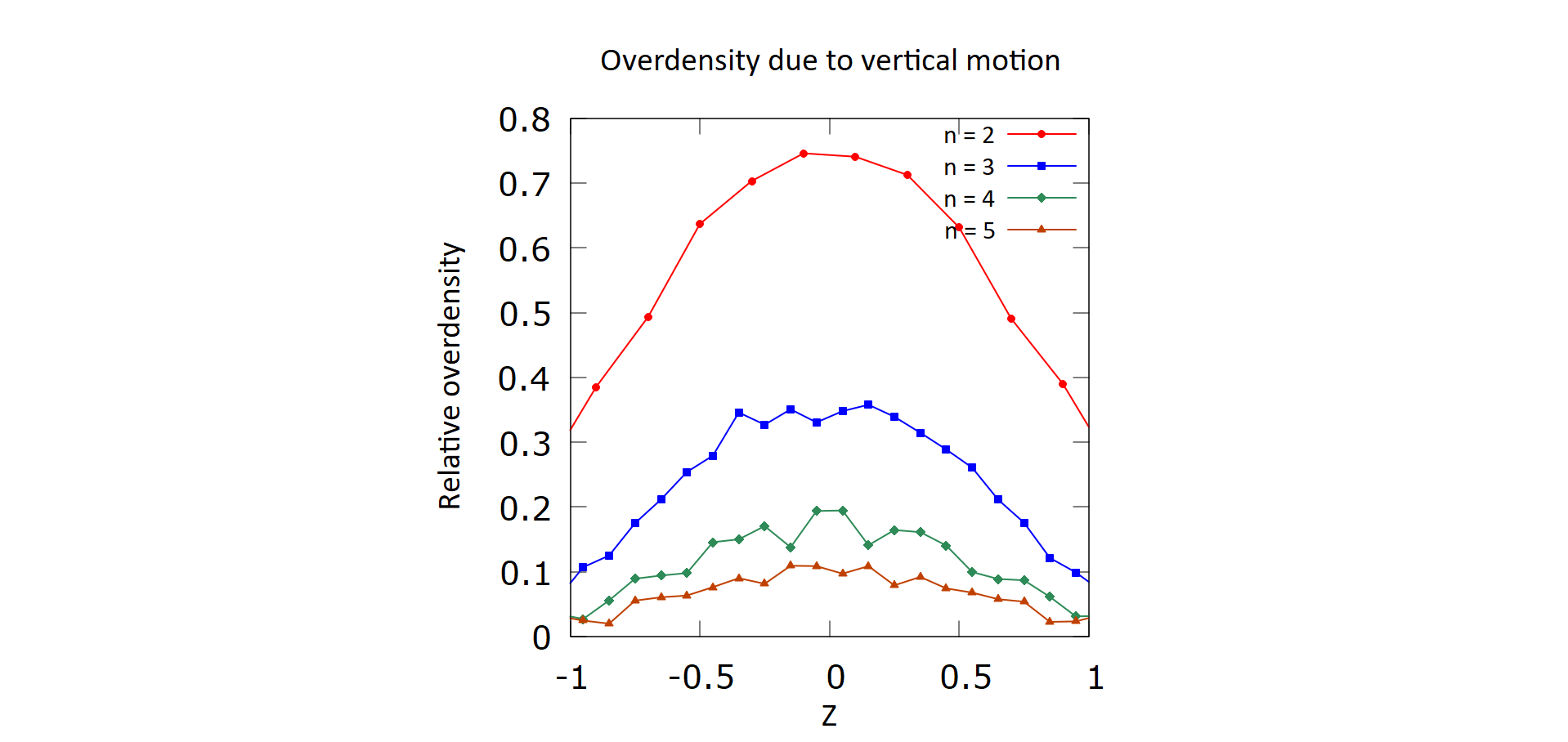}
\caption{\label{overdensityZ}Relative overdensities of the stars (same as Fig.~\ref{overdensityX}) due to vertical motion as functions of the rescaled vertical coordinate $Z = \frac{z}{p}$.}
\end{figure}

Next, we study how the bulk properties of a distribution of stars are affected by the passing of a caustic ring. We find that the induced bulk velocities of the stars cannot be more than a few km/s. The star overdensity around a caustic ring depends upon the velocity dispersion of the stars. Since a caustic ring with smaller $n$ has stronger gravitational field and is surrounded by stars with lower velocity dispersion, it causes a larger stellar overdensity. Figures \ref{overdensityX}-\ref{overdensityZ} show the overdensities of stars around different caustic rings due to radial and vertical dynamics. The maximum total overdensity is approximately $120\%$ around the $n=2$ caustic ring at $20$ kpc which supports the claim of Ref.~\cite{Natarajan07} that the Monoceros Ring \cite{Newberg02, *Yanny03, *Ibata03, *Rocha03} may be caused by the second caustic ring. The total overdensity of $45\%$ near the $n = 3$ caustic ring may explain the existence of the Binney and Dehnen ring at $13.6$ kpc \cite{Binney97}. The overdensities are smaller for larger $n$, e.g. approximately $30\%$ and $15\%$ for the $n= 4$ and $5$ caustic rings, respectively. These overdensities, within a few kpc from us, may be observable in upcoming astronomical data.

\begin{figure}
\begin{center}
\includegraphics[scale=0.5]{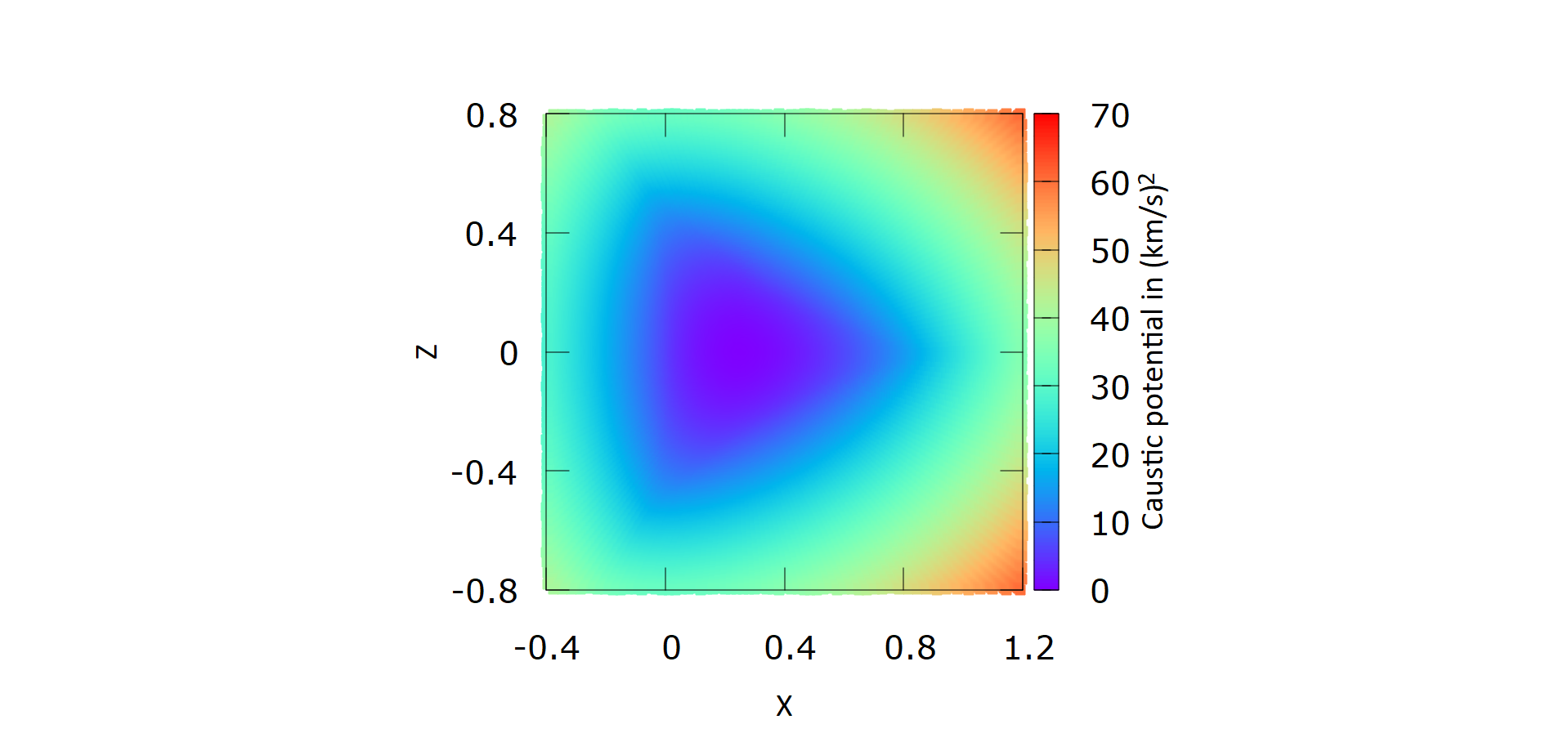}
\caption{\label{causticpot}Two-dimensional plot of the gravitational potential $\Phi_c (X, Z)$ of a caustic ring as a function of the rescaled coordinates $X = \frac{\rho - a}{p}$ and $Z = \frac{z}{p}$. The parameters of the caustic ring are the same as those in Fig.~\ref{gplot}. The center of the tricusp ($X_0 = 0.25$, $Z_0 = 0$) is chosen as the reference point. The potential has a triangular shape inherited from the tricusp.}
\end{center}
\end{figure}

\begin{figure}
\includegraphics[scale=0.5]{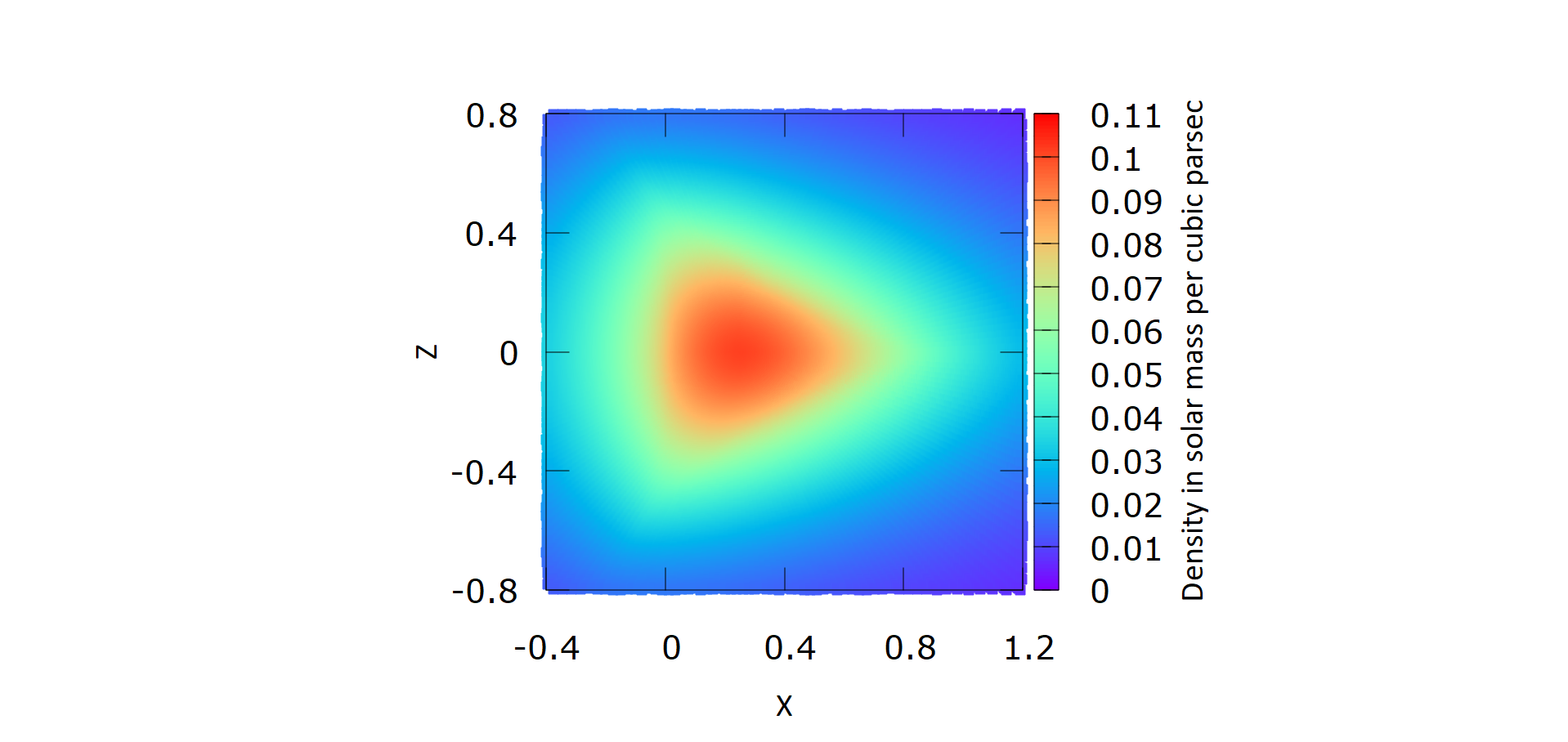}
\caption{\label{densityplot}Two-dimensional plot of the density of gas $d_g (X, Z)$ (Eqs.~\ref{rhoxz}, \ref{phixzn}) when the gas is in thermal equilibrium in the gravitational potential of the caustic (shown in Fig.~\ref{causticpot}) and of the gas itself. The velocity dispersion $\sigma_g$ of the gas is chosen to be $5$ km/s. The density profile has a triangular shape reflecting that of the tricusp.}
\end{figure}

We also study the distribution of interstellar gas near the $n=5$ caustic ring which is the one closest to us. The gas dynamics is much faster than the time scale over which the radius of a caustic ring changes. We consider the caustic ring to be static and assume the gas to be in thermal equilibrium in the gravitational potential of the caustic (shown in Fig.~\ref{causticpot}) and of the gas itself. Using a Green's function, we iteratively solve the Poisson's equation to find the density of gas. The density profile of gas in $\rho$-$z$ plane shown in Fig.~\ref{densityplot} has a triangular shape as a result of the tricusp cross section of the caustic ring. In the recent GAIA sky map \cite{GAIADR1, *GAIADR2}, two triangular features in both tangent directions to the fifth caustic ring are observed \cite{Chakrabarty18}. However, the observed features have sharper edges than the features (Fig.~\ref{densityplot}) we obtain.

\raggedbottom

\begin{acknowledgments}
The authors are grateful to Heywood Tam for the code for calculating the gravitational field of a caustic ring. This work was supported in part by the U.S. Department of Energy under Grant No. DE-FG02-97ER41209 and by the Heising-Simons Foundation under Grant No. 2015-109.
\end{acknowledgments}

\bibliography{caustic_bib}

%\clearpage

\end{document}